\documentclass[twocolumn,amsmath,amssymb,floatfix,prb,preprintnumbers,footinbib, showpacs]{revtex4-1}
\newlength{\stdwidth}
\usepackage[dvips]{graphicx}
\usepackage{bbm}
\usepackage{amsfonts}
\usepackage{amsmath}
\usepackage{graphicx}
\usepackage[utf8]{inputenc}

\usepackage[colorlinks=true,citecolor=Blue,linkcolor=Blue, breaklinks=true]{hyperref}

\newcommand{\subref}[2]{\ref{#1}{\color{Blue}({#2})}}
\newcommand{\sub}[1]{{#1}}

\usepackage[pdftex]{color}
\definecolor{contentlistcolor}{rgb}{.5,.5,.7}
\definecolor{Blue}{rgb}{.0,.0,.8}

\def\mathvecfont#1{\textbf{\em #1}}
\newcommand{\myvec}[1]{\mathvecfont{#1}}

\newcommand{\ci}{\mathrm{i}}
\newcommand{\e}{\mathrm{e}}
\newcommand{\eref}[1]{(\ref{#1})}

\begin{document}
\title{Bloch--Zener Oscillations in Graphene and Topological Insulators}

\author{Viktor Krueckl}
\affiliation{Institut f\"ur Theoretische Physik, Universit\"at Regensburg, D-93040 Regensburg, Germany}

\author{Klaus Richter}
\affiliation{Institut f\"ur Theoretische Physik, Universit\"at Regensburg, D-93040 Regensburg, Germany}

\date{\today}

\begin{abstract}
We show that superlattices based on zero-gap semiconductors such as graphene and mercury telluride exhibit
characteristic Bloch--Zener oscillations that emerge from the coherent superposition of Bloch oscillations
and multiple Zener tunneling between the electron and hole branch.
We demonstrate this mechanism by means of wave packet dynamics in various spatially periodically modulated
nanoribbons subject to an external bias field.
The associated Bloch frequencies exhibit a peculiar periodic bias dependence which we explain within a
two-band model.
Supported by extensive numerical transport calculations, we show that this effect gives rise to distinct
current oscillations observable in the $I$-$V$ characteristics of graphene and mercury telluride
superlattices.

\end{abstract}



\maketitle

\section{Introduction}
Bloch oscillations, the periodic motion of particles in a superlattice subject to a constant external field,
represent a fundamental phenomenon in transport through periodic potentials.
Predicted already in the early days of quantum mechanics~\cite{Bloch1929, Zener1934}, Bloch oscillations have
been observed in various fields of physics, ranging from earlier experiments in semiconductor
superlattices~\cite{Feldmann1992, Leo1992, Waschke1993} via  cold atoms in optical
lattices~\cite{BenDahan1996, Wilkinson1996} to classical optical~\cite{Pertsch1999, Morandotti1999} and
acoustic~\cite{Sanchis-Alepuz2007} waves.
While many aspects of conventional Bloch oscillations can be explained by a single band description,
particularly interesting effects arise in the case of two coupled minibands~\cite{Fukuyama1973}
energetically separated from further bands. 
Then partial Zener tunneling at avoided crossings of the two minibands can lead to a coherent superposition 
of Bloch oscillations~\cite{Rotvig1995, Hone1996}, {\em i.e.} to a splitting, followed by as subsequent
recombination of a Bloch oscillating wave packet.
This gives rise to a variety of Rabi-type interference phenomena, in particular  double-periodic motions
coined Bloch--Zener (BZ) oscillations~\cite{Breid2006, Breid2007, Abumov2007}.
Signatures of this effect have already been detected in the THz emission of AlGaAs
superlatices~\cite{Shimada2004}, and even the population dynamics have been measured recently for
light~\cite{Dreisow2009} and atomic matter waves~\cite{Kling2010} in especially tailored binary lattices.

However, materials with a linear Dirac spectrum~\cite{Wallace1947} naturally serve the effect, since only a
small gap is opened by a spatially periodic modulation allowing for Zener tunneling between electron and
hole states.
Such materials are now at hand with the discovery of graphene~\cite{Novoselov2004, Zhang2005} 
and the advent of topological insulators~\cite{Kane2005, Kane2005_2, Bernevig2006, Bernevig2006_2}
first realized in two-dimensional mercury teluride (HgTe) heterostructures~\cite{Koenig2007, Roth2009}.
Interesting phenomena for graphene based periodic superstructures have already been
theoretically predicted like the formation of extra Dirac cones~\cite{Park2008, Barbier2010, Brey2009}
and the appearance of a negative differential conductance~\cite{Ferreira2011}.
Furthermore, recent experiments have realized graphene superlattices with periodicities down to
a few nm~\cite{Meyer2008}.

This raises the question for the existence of peculiarities of Bloch oscillations
in graphene and topological insulator superlattices that we address 
in this manuscript.
We are not aware of work showing unconventional features in 
graphene-based Bloch oscillations.
Up to now, only the semiclassical approach was adapted to a linear dispersion~\cite{Dragoman2008}
and, without reference to Bloch oscillations numerical evidence for a negative
differential conductance was reported~\cite{Ferreira2011}.
We show that besides conventional Bloch oscillations, multiple Zener tunneling between the
coupled electron and hole branches leads to distinct BZ 
oscillations that appear to be 
naturally present in superlattices made of systems with Dirac-like dispersion.

This paper is structured as follows:
In Section~\ref{sec::GrapheneWP} we show the influence of BZ
oscillations on the wave packet motion in a graphene nanoribbon and the
influence on the frequency spectrum.
Subsequently, we introduce in Section~\ref{sec::Model} a two-band model to explain the effect in the
frequency spectrum and the influence of BZ oscillations on the electron-hole polarization.
In Section~\ref{sec::GrapheneT} we show, that the occurrence of BZ oscillations
can be seen as distinct features in the current through graphene nanoribbons.
In Section~\ref{sec::HgTe} we present results that feature the special frequency pattern of BZ
oscillations, as well as their signatures in transport, in nanoribbons made of the
topological insulator mercury telluride.

\section{Wave-packet motion in graphene superlattices}
\label{sec::GrapheneWP}
%
%
\begin{figure}[tb]
\centering
\includegraphics[width=\stdwidth]{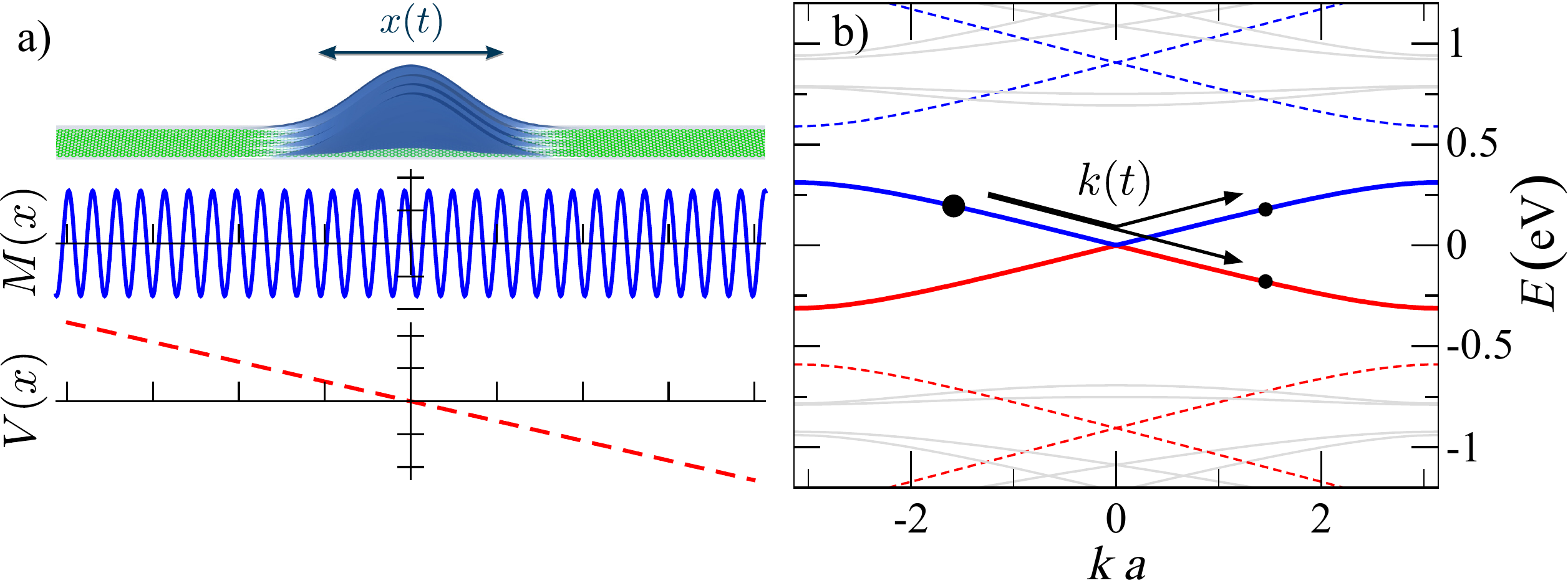}
\caption{(\emph{Color online})
Exemplary setup for Bloch--Zener oscillations in a graphene nanoribbon. 
\sub{a}) Sketch of a Gaussian wave packet in presence of a periodic mass potential $M(x)=M_0 \sin(2 \pi x/a)$ and
an electrostatic drift potential $V(x)= - e E_D x$.
\sub{b}) Bandstructure of the superlattice with small avoided crossing at $k=0$
(nanoribbon width $W = 10~a_0$, $a = 10\sqrt{3}~a_0$, $M_0 = 0.1~t$).
Thick and dashed lines show the first and second Bloch band from the metallic
armchair mode. The gray lines represent higher transversal modes.
}
\label{figsketch}
\end{figure}
%
%
An insight into the dynamics of BZ oscillations  can be gained 
by the time-evolution of a wave packet on a graphene nanoribbon in presence
of a periodic mass potential $M(x)$ and a linear electrostatic drift
potential $V(x)$ as sketched in Fig.~\subref{figsketch}{a}.
To this end we model the electronic structure of graphene by a conventional
tight-binding Hamiltonian~\cite{Nakada1996}
%
%
\begin{equation}
H_\mathrm{tb} =
    \sum_{\langle i~j\rangle, \beta} t~c_{i,-\beta}^\dagger c_{j,\beta} 
   + V~c_{i, \beta}^\dagger c_{i, \beta}
   + M~\beta~c_{i, \beta}^\dagger c_{i, \beta}  
\label{tbHamiltonian}
\end{equation}
%
%
where $\langle i~j\rangle$ denotes neighbouring unit cells and $\beta = \pm1$ the sublattice degree of freedom.
The initial wave packet is created by diagonalizing the periodic Hamiltonian $H(k)$  of one unit-cell of the
infinite ribbon.
By means of the transversal eigenfunctions $\chi_n(y,k)$ we create an initial electron-like
wave packet
%
%
\begin{equation}
\psi_n(x,y) = \int_{-\infty}^{\infty} \chi_n(y,k)~\e^{\ci k x}~\e^{-\frac{1}{2} k^2 \delta^2} \mathrm{d}k
\label{initialWP}
\end{equation}
%
%
%
with a Gaussian broadening $\delta$.
Since the armchair boundary mixes the two graphene valleys the wavefunction comprises
several nodes in lateral direction.
The time-evolution is calculated by an expansion of the time-evolution operator in
Chebychev polynomials~\cite{krueckl2009}.
In presence of a periodic mass potential
%
%
\begin{equation}
M(x) = M_0\sin( 2 \pi x / a) ,
\label{masspotential}
\end{equation}
%
%
%
where $M_0$ is the strength of the periodic mass, and $a$ the periodicity length,
the bandstructure of the superlattice exhibits a small anti-crossing at $k=0$ and
a large bandgap between the first and the higher Bloch bands as shown
in Fig.~\subref{figsketch}{b}.
In presence of the a linear drift potential 
%
%
\begin{equation}
V(x) = - e E_D x \text{,}
\label{driftpotential}
\end{equation}
%
%
with $E_D$ as the strength of the drift field, the wave packet starts to accelerate.
Because of its extent in the longitudinal direction, the wave packet is localized
in momentum space with a distinct average momentum $k(t)$ in $x$-direction.
Given the periodicity of the bandstructure,
a sawtooth behavior of $k(t)$ is obtained known as Bloch oscillations.

\begin{figure}[tb]
\centering
\includegraphics[width=\stdwidth]{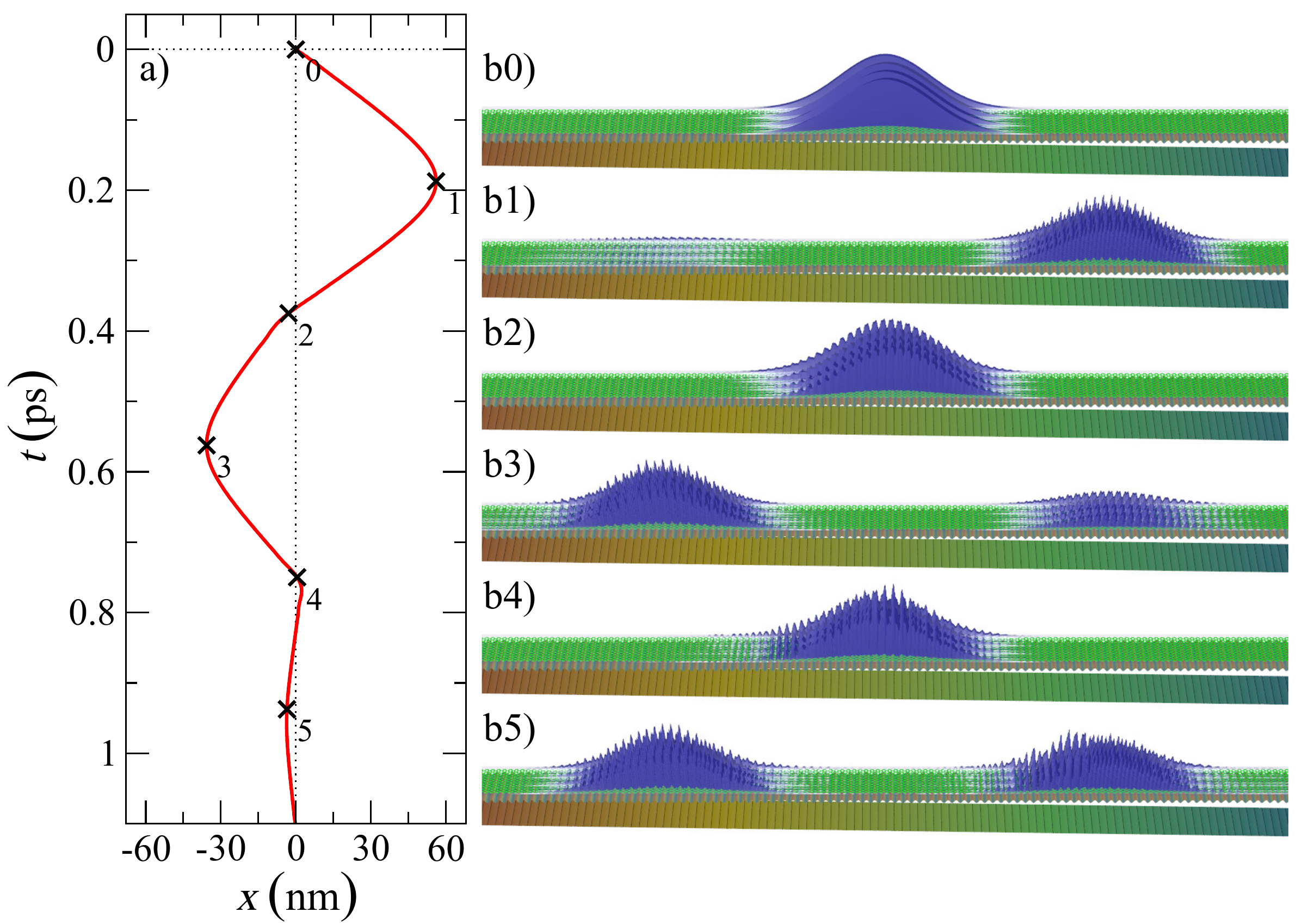}
\caption{(\emph{Color online})
Snapshots of a wave packet in the course of Bloch--Zener oscillations.
\sub{a}) Center-of-mass motion of a wavepacket on a graphene nanoribbon ($W=10a_0$, $a=10\sqrt{3} a_0$, $M_0=0.1 t$).
\sub{b}) Snapshots of the probability distribution of the wavepacket for the corresponding times marked with crosses
in panel (a). Please note the video of the dynamics in the online version of the Supplemental Material~\cite{Support}.
}
\label{figtraj}
\end{figure}

However, the dynamics in a graphene nanoribbon shows additional features due to the strong
coupling between electron and hole states.
Therefore we study a typical trajectory
%
%
\begin{equation}
x(t) = \langle \psi(t) \vert \hat x \vert \psi(t) \rangle
\end{equation}
%
%
%
of the center-of-mass (COM) as shown in Fig.~\subref{figtraj}{a}.
%
%
%
%
Initially, the wave packet is chosen electron like and a snapshot of the
probability distribution is shown in Fig.~\subref{figtraj}{b0}.
During the first Bloch cycle the probability distribution is predominantly to the
right of the initial position [see Fig.~\subref{figtraj}{b1}].
This region features a negative electrostatic potential and accordingly the part of
the wave packet with electron character performs Bloch oscillations in this region.
In Fig.~\subref{figtraj}{b2} the electron and hole parts meet again in momentum space
and as a consequence tunneling from the electron to the hole branch is possible
as sketched by the bullets in Fig.~\subref{figsketch}{b}.
As a result the hole-like part of the wave-packet increases and in subsequent
time steps the COM trajectory reaches negative values.
The corresponding snapshot at the turning point of the the wavefunction in 
Fig.~\subref{figtraj}{b3} shows a big hole-like state on the 
left side and a smaller electron-like state on the right side.
After the next tunneling the probability distribution between electron and hole states 
is almost equal, thus the COM motion is strongly suppressed.
Because of the periodic mass potential, the gaps between the first Bloch band 
and higher bands is bigger than the gap between electron and hole states as
shown in Fig.~\subref{figsketch}{b}.
As a result the tunneling into higher bands is very unlikely and there is no
damping of the oscillations due to leakage into higher bands.

%
%
%
\begin{figure}[t]
\centering
\includegraphics[width=\stdwidth]{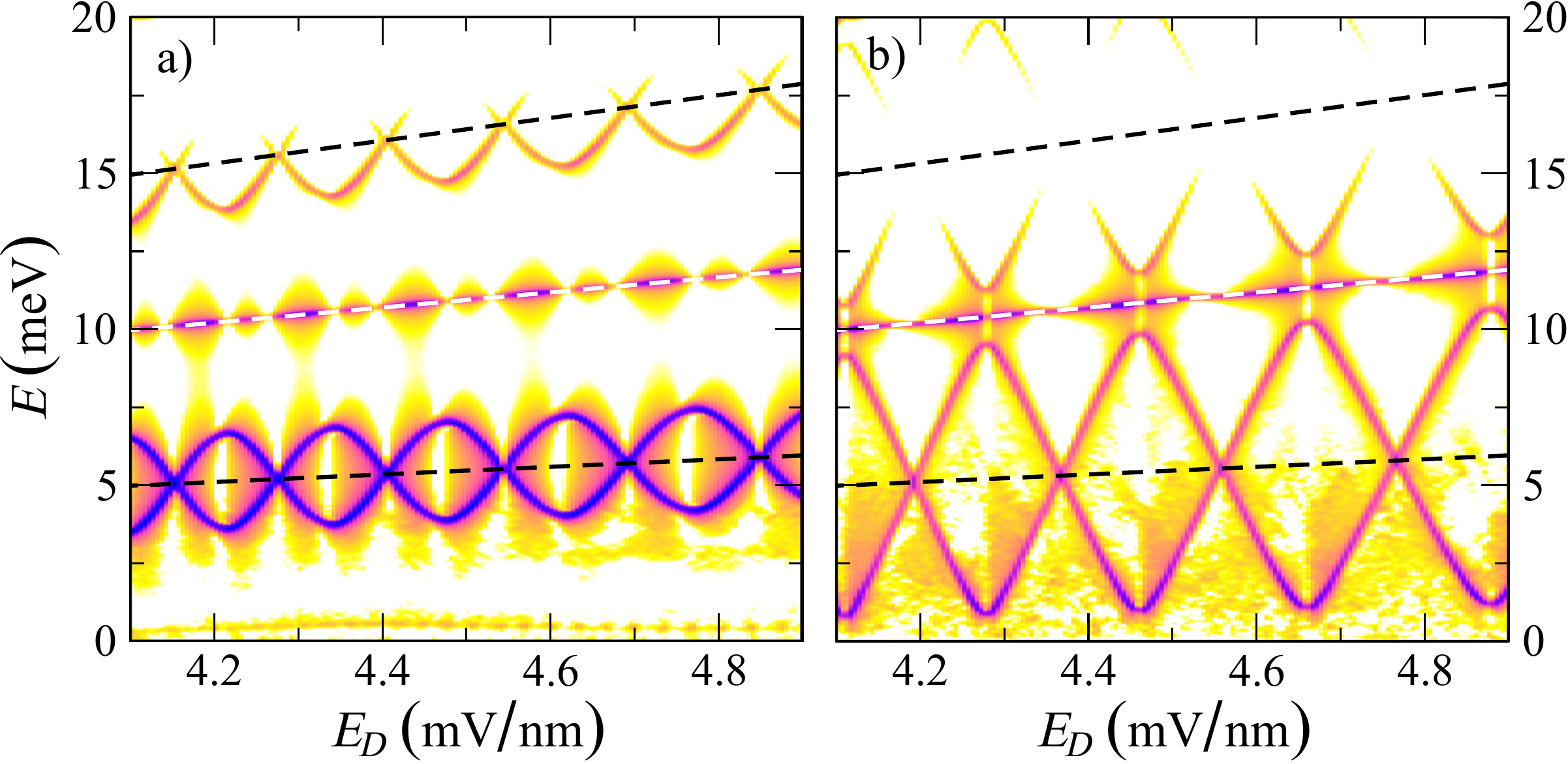}
\caption{(\emph{Color online})
Frequency spectra $E=\hbar \omega$ from the center-of-mass motion of a wave packet for varying drift field $E_D$
for (a) moderate ($M_0=0.1t$) and (b) stronger periodic potential ($M_0=0.2t$).
Dark colors represent strong intensities.
The dashed lines correspond to \{1/2, 1, 3/2\} times the conventional Bloch frequency.
}
\label{figwpgraphene}
\end{figure}
%
%

To study the dynamics of the tunneling between the electron and the hole branch
in more detail we perform a frequency analysis of the COM motion for different 
fields $E_D$.
The Fourier amplitudes of the dominant frequency contributions are visualized by dark colors 
in Fig.~\ref{figwpgraphene}.
Besides the conventional Bloch frequency marked by a white dashed line,
the resulting spectrum shows a pronounced interweaving pattern around half of this frequency (black dashed line).
A stronger periodic potential, and thereby an increased gap between electron and hole branch, leads to a
rhombic structure as shown in Fig.~\subref{figwpgraphene}{b}.
These periodic features in the frequency spectrum arise from the interplay between Bloch oscillations
and the splitting of the wave packet into the electron  and hole branches at $k(t)=0$ (see Fig.~\subref{figsketch}{a}).
The persistent sequence of tunneling events between the two branches and the subsequent
interference leads to a new set of frequencies which can be unterstood by means of
the following model.

\section{Analytical model for BZ oscillations}
\label{sec::Model}
\begin{figure}[tb]
\centering
\includegraphics[width=\stdwidth]{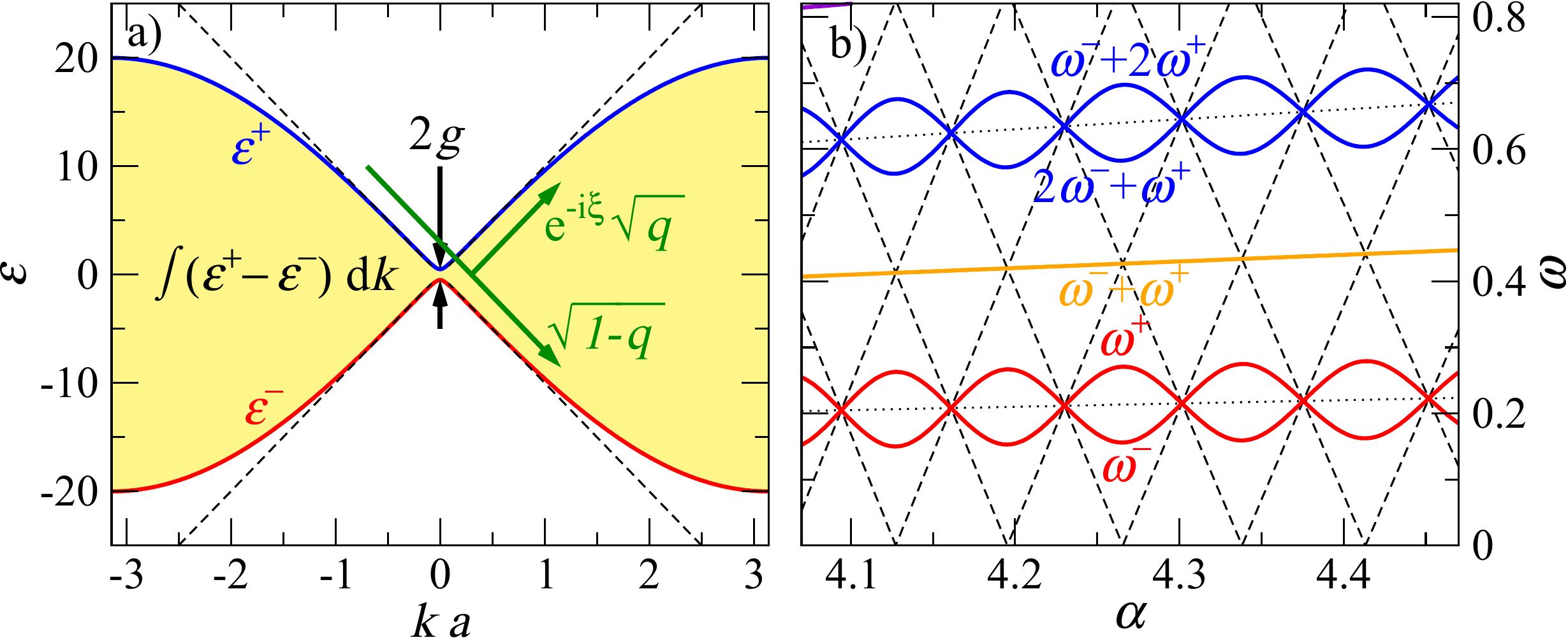}
\caption{(\emph{Color online})
\sub{a}) 
Bandstructure of the Dirac model Hamiltonian~(\ref{modelH}) for $v = 1$, $\hbar = 1$, $a = 1/10$, $g = 1/2$.
The shaded (yellow) area denotes the integral~\eref{eqmomentumarea}. 
\sub{b})
Frequency spectrum of the Bloch oscillations for different drift accelerations $\alpha=e E_D /\hbar$.
Solid lines show the frequencies $n \omega^+ + m \omega^-$ given by Eq.~(\ref{modelFreq}), dotted (dashed)
lines the strong (weak) tunneling limit.
}
\label{figmodel}
\end{figure}

In the following we quantitatively explain these characteristic BZ features using a periodically modulated
one-dimensional Dirac model Hamiltonian,
%
%
\begin{equation}
H(t) = \tfrac{2 \hbar v}{a} \sin{\left(\tfrac{a k(t)}{2}\right)}\,\sigma_z + g\,\sigma_x \text{.}
\label{modelH}
\end{equation}
%
%
Here $a$ is the period, $v$ is the Fermi velocity  and $g$ the energy gap between the electron and the
hole states.
The resulting bandstructure is given by
%
%
\begin{equation}
\epsilon^\pm(t) = \pm \sqrt{g^2 + 2 ( \hbar v / a )^2 \left[1- \cos(a k) \right]}
\end{equation}
%
%
as shown in Fig.~\subref{figmodel}{a}.
A comparison with the full tight-binding calculation of the graphene nanoribbon 
in Fig.~\subref{figsketch}{b} shows a very good correspondence.
%
%
%
%
An external electric drift field $E_D$ enters the equations of motion for the quasi-momentum $k(t)$ as
%
%
$
\hbar\,\partial_t k(t) = e E_D
$
%
%
leading to a time evolution of $k(t) = \alpha t$  linear in $t$ where  $\alpha = {e E_D}/{\hbar}$.
Conventional Bloch oscillations with frequency $\omega_B=\alpha a$ arise from the
periodicity of $k(t)$ in momentum space in the interval $[-\frac{\pi}{a},\frac{\pi}{a}]$.
The phase $\phi$ between the two branches accumulated during one oscillation is given by a free propagation
and thus 
%
%
\begin{equation}
\phi = \frac{\mathcal A}{e E_D}  \approx \frac{16 v}{a^2 \alpha}
\label{modelphase}
\end{equation}
%
%
with
%
%
\begin{equation}
\mathcal A = \int_{-\pi/a}^{\pi/a} (\epsilon^+ - \epsilon^-) \mathrm{d}k
\label{eqmomentumarea}
\end{equation}
%
%
the area in momentum space as
depicted in Fig.~\subref{figmodel}{a}.
This free propagation can be expressed by the matrix
%
%
\begin{equation}
U_0 = 
\begin{pmatrix}
\e^{\ci \phi/2} & 0 \\
0 & \e^{-\ci \phi/2}
\end{pmatrix}\text{.}
\end{equation} 
%
%
Additional to conventional Bloch oscillations on either branch, there is a strong periodic tunneling between
the electron and the hole states close to the anti-crossing at $k=0$.
There, the Hamiltonian~(\ref{modelH}) can be linearized [dashed lines in Fig.~\subref{figmodel}{a}], leading
to a typical Landau--Zener tunneling problem~\cite{Landau1932, Zener1932, Stueckelberg1932}:
%
%
\begin{equation}
H_\mathrm{LZ} =
\begin{pmatrix}
\hbar v \ \alpha t & g \\
g & -\hbar v \ \alpha t
\end{pmatrix}\text{.}
\end{equation}
%
%
The scattering between the different branches is described by
%
%
\begin{equation}
S_0 = 
\begin{pmatrix}
\e^{- \ci \xi} \sqrt{q} & \sqrt{1-q} \\
\sqrt{1-q} & -\e^{\ci \xi} \sqrt{q}
\end{pmatrix}
\end{equation}
%
%
with the tunneling rate 
$q = 1-\e^{- 2 \pi \delta}$, $\delta = \frac{g^2}{2 \hbar^2 v \alpha}$, and
$\xi = \frac{\pi}{4} + \arg(1 - \ci \delta) + \delta(\log \delta-1)$ is an additional tunneling phase.
From this we can deduce the scattering matrix
%
%
\begin{equation}
S = 
\begin{pmatrix}
\e^{\ci (\phi/2-\xi)} \sqrt{q} & \sqrt{1-q} \\
\sqrt{1-q} & -\e^{\ci (\xi-\phi/2)} \sqrt{q}
\end{pmatrix} \text{,}
\end{equation}
%
%
which describes the time-evolution of the electron and hole branch for one Bloch cycle.
Using this matrix we derive scattering eigenstates
%
%
\begin{equation}
\chi^\pm = \frac{1}{\sqrt{\mathcal N}}
\begin{pmatrix}
  \sqrt{q} \cos(\phi/2-\xi) \pm \sqrt{1 - q \sin^2(\phi/2-\xi) }    \\ \sqrt{1-q}~\e^{\ci \phi/2}
\end{pmatrix}
\end{equation}
%
%
with the corresponding eigenvalues $\e^{\ci \beta^\pm}$ where
%
%
\begin{equation}
\beta^\pm = 
\arccos\left( \pm \sqrt{1- q \sin^2(\phi/2 - \xi) } \right) \text{.}
\end{equation}
%
%
The phases $\beta^\pm$ of the scattering eigenstates depend periodically on the
phase difference $\phi$ between electron and hole branch.
This periodicity leads to two new Bloch frequencies
%
%
\begin{equation}
\omega^\pm = 
\frac{\alpha a}{\pi} 
  \arccos\big [
    \pm \sqrt{q} \sin(\phi/2 - \xi)
  \big ]
\text{.}
\label{modelFreq}
\end{equation}
%
%
Unlike conventional Bloch oscillations these frequencies do not simply depend linearly on the drift strength
$\alpha$, but show a rapid interweaving pattern strongly changing with $\alpha$, as shown in
Fig.~\subref{figmodel}{b}, owing to coherences from combined dynamics on the hole and electron branch.
The limiting cases can be understood as follows.
For strong coupling, the tunneling rate $q \rightarrow 0$ leads to a frequency
$\omega^\pm \rightarrow \omega_B/2$ [dotted line in Fig.~\subref{figmodel}{b}], since for
every Bloch cycle the states tunnel completely between the two branches in momentum space and hence the
complete cycle in position space is twice as long.
In the opposite, weak coupling limit $\omega^\pm \rightarrow a \alpha [1/2 \pm (\phi/2-\xi)/\pi ] \mod 1 $
leading to a rhombic frequency pattern shown as dashed lines in Fig.~\subref{figmodel}{b}.
For intermediate tunneling rates the frequencies show a smooth transition between these limiting cases and
are in very good agreement with the numerically calculated spectra of Fig.~\subref{figwpgraphene}{a,b}.

%
%
%
\begin{figure}[t]
\centering
\includegraphics[width=\stdwidth]{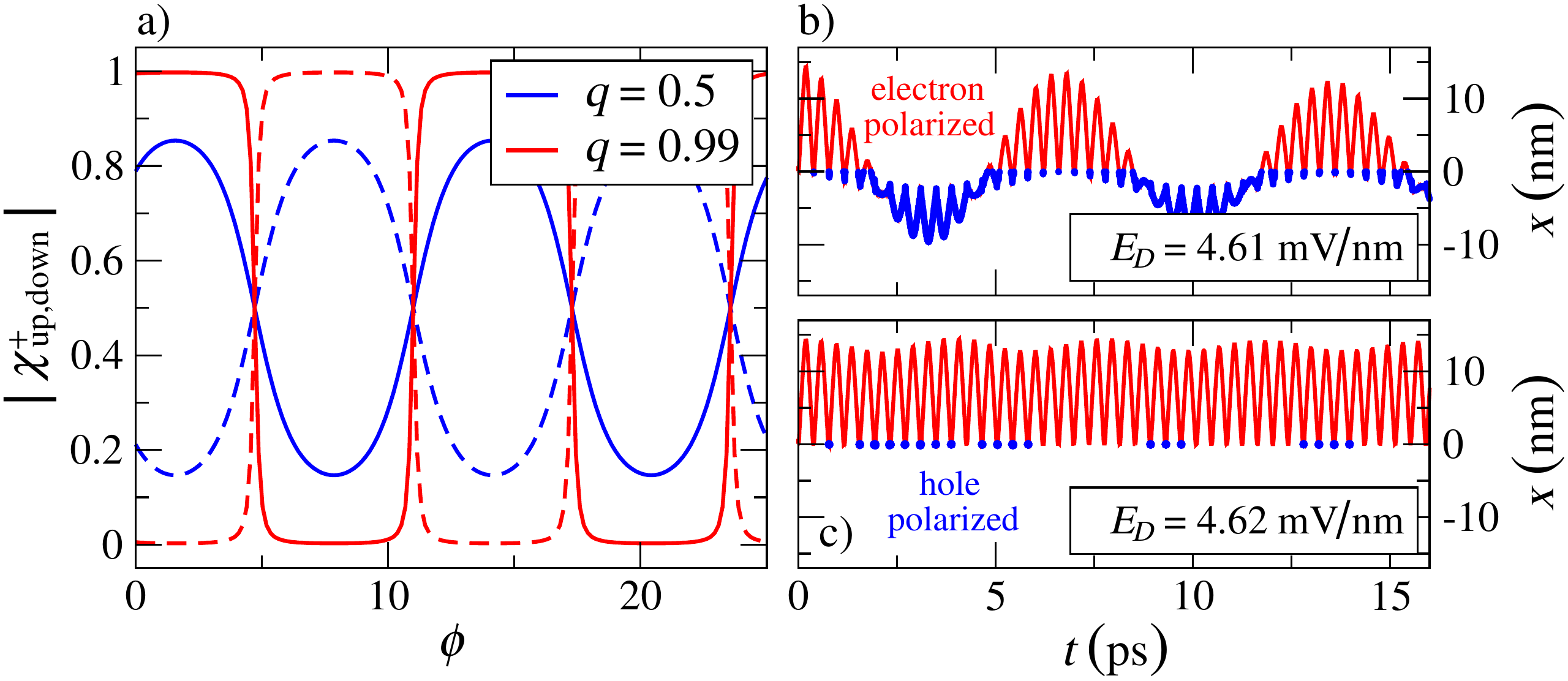}
\caption{(\emph{Color online})
\sub{a}) 
Polarization dependence on the phase difference between electron and hole branch
of the scattering eigenstate $\chi^+$
(solid line shows the upper spinor entry, dashed line the lower spinor entry). 
Panels (\sub{b}), (\sub{c})  show the center-of-mass motion of an initially
electron-polarized wavepacket on a graphene nanoribbon superlattice $M(x)=M_0 + V(x)$
for different drift fields $E_D$
($M_0= 50\,\mathrm{meV}$, $V(x)= 300\,\mathrm{meV}\sin(2 \pi x/a)$, $a=10\sqrt{3}a_0$).
Blue dots depict regions with a negative amplitude, corresponding to a
wavepacket with strong hole character.
}
\label{figpol}
\end{figure}
%
%
Furthermore, the scattering eigenstates show a strong polarization
dependence (electron or hole type character) on the phase $\phi$, Eq~\eref{modelphase}.
If the one-dimensional model Hamiltonian~\eref{modelH} is considered for
$g^2 \gg 2 \hbar^2 v \alpha$, the tunneling rate $q \rightarrow 1$,
which results in strongly electron or hole polarized states $\chi^\pm$
for almost all values of $\phi$ .
The absolute value of the spinor entries is always very close to 
one or zero as shown in Fig.~\subref{figpol}{a}.
Nevertheless, the polarization breaks down whenever the difference between the 
phase of the electron and hole branch is
%
%
\begin{equation}
\phi = 2(n \pi + \xi)+\pi
\label{phasecond}
\end{equation}
%
%
where $n \in \mathbb N$.
This alternating weight of the spinor between the electron and hole type states 
for different drift fields $E_D$ can be also deduced from the COM motion of wave packets
with fixed initial polarization.
If the drift field is adjusted such that the phase condition~\eref{phasecond}
is approximately satisfied,
the COM motion of the initially electron-like configuration
exhibits oscillations ranging from
$-15\,\mathrm{nm}$ to $15\,\mathrm{nm}$ for $E_D=4.61\,\mathrm{mV}/\mathrm{nm}$
as shown in Fig.~\subref{figpol}{b}.
Since conventional Bloch oscillations in a single band are restricted to positive or negative
values the trajectories imply strong tunneling between the electron and hole states.
For values of $E_D$ where condition~\eref{phasecond} is not fulfilled,
\emph{e.g.} $E_D=4.62\,\mathrm{mV}/\mathrm{nm}$ in Fig.~\subref{figpol}{c},
the trajectories of the different polarizations do not significantly cross the origin, thus they preserve
their electron-hole character.
As a consequence, if charge transport through a system comprises a transition
from electron to hole states the current should strongly depend on the BZ oscillations
within the superstructure.

\section{Transport in graphene-based superlattices}
\label{sec::GrapheneT}
In the following, we consider charge transport through graphene nanoribbon
based superlattices and demonstrate that BZ
oscillations lead to clear-cut features in the $I$-$V$ characteristics.
%
%
\begin{figure}[tb]
\centering
\includegraphics[width=\stdwidth]{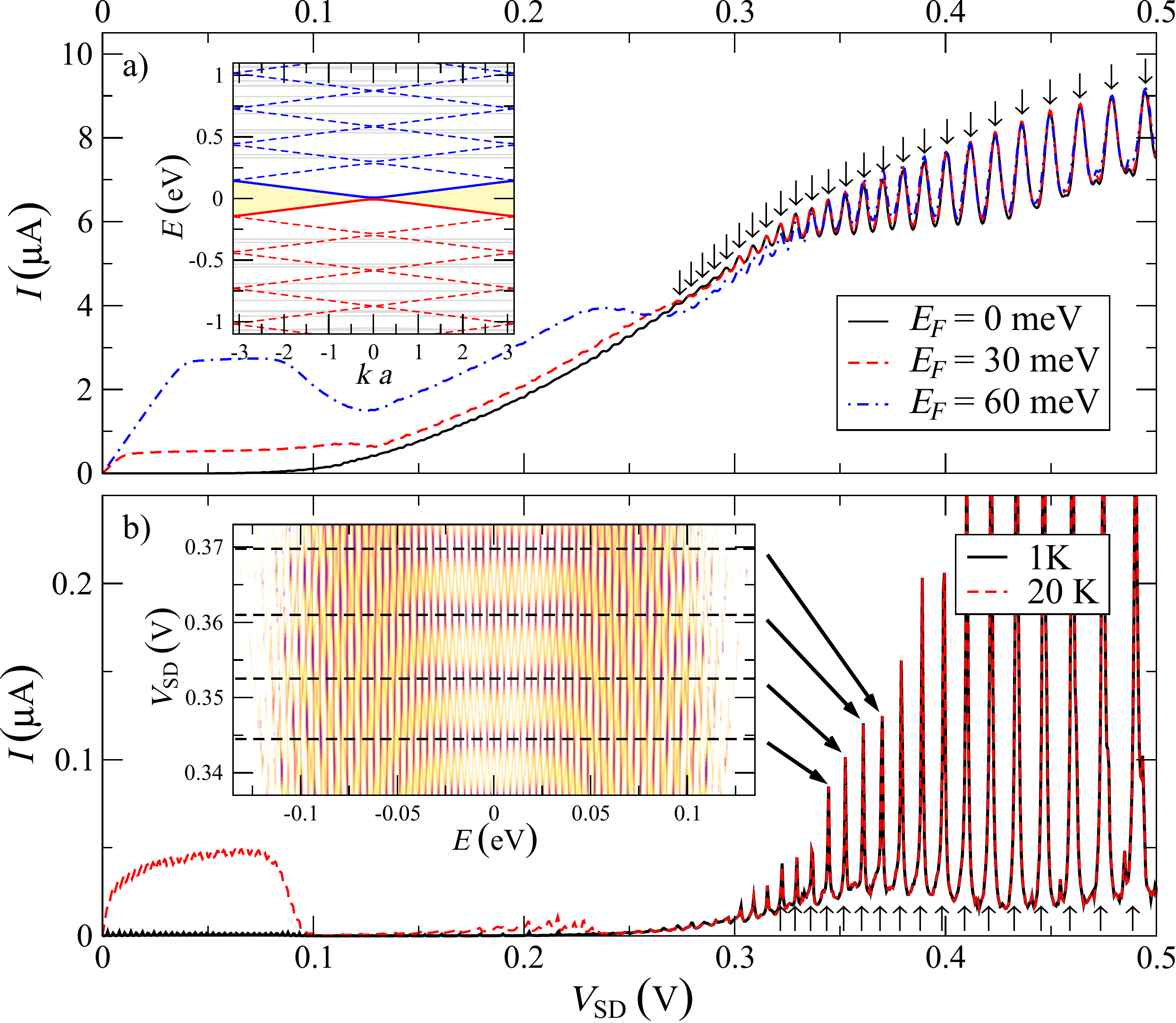}
\caption{(\emph{Color online})
Current-voltage characteristics for graphene nanoribbon superlattices ($L=3000 \sqrt{3} a_0$, $W=10 a_0$,
$a=30 \sqrt{3} a_0$, $V_0=500~\mathrm{meV}$) for 
(a)    different Fermi energies ($M_0=20\mathrm{meV}$, $T=20\mathrm{K}$) and
(b)    different temperatures ($M_0=50\mathrm{meV}$, $E_F=0$)
showing pronounced signatures of Bloch--Zener oscillations at higher bias.
Arrows mark expected peak positions from phase condition~\eref{phasecond}.
Upper inset: Bandstructure (for $M_0=20\mathrm{meV}$),
lower inset: Transmission map $T(E, V_\mathrm{SD})$ used in Eq.~(\ref{currentEq}) to get the current
of panel (b); dark colors represent high transmissions.
\label{figGrT}
}
\end{figure}
%
%
To this end we model a graphene nanoribbon of width $W$ and length $L$ by the tight-binding Hamiltoninan of
Eq.~(\ref{tbHamiltonian}), now with a periodic electrostatic potential $V_0\sin(2 \pi x / a )$ leading to a
superlattice mini-bandstructure as shown in the inset of Fig.~\subref{figGrT}{a}.
A small constant mass term $M(x)=M_0$ is additionally considered which opens up a gap commonly present in
experiments on graphene nanoribbons~\cite{Han2010}.
We assume a linear potential drop $e V_\mathrm{SD} x/L$ due to the source-drain voltage $V_\mathrm{SD}$
between the graphene leads at $x=\pm L/2$.
The current is calculated by means of the Landauer-B\"uttiker formalism~\cite{Buettiker1985},
%
%
\begin{equation}
I(V_\mathrm{SD}) = 
   \frac{2e}{h} \int_{-\infty}^\infty  T(E, V_\mathrm{SD}) [f^+(E) - f^-(E) ]  \mathrm{d}E \text{,}
\label{currentEq}
\end{equation}
%
%
with the Fermi functions $f^\pm(E) = \{ 1+ \exp[(E \mp V_\mathrm{SD}/2)/k_\mathrm{B}T] \}^{-1}$.

As shown in Fig.~\ref{figGrT}, the current through the nanoribbon is governed by a conventional increase
with the bias window for small $V_\mathrm{SD}$, followed by a region of negative differential 
conductance typical for superlattices. 
At higher bias, $V_\mathrm{SD} > 0.3\,\mathrm{V}$, we observe the emergence of distinct current oscillations
that get more pronounced with increasing gap size, see Fig~\subref{figGrT}{b}.
Due to the bias between source and drain electrode the particles traversing the superlattice must change
their electron-hole character.
However, states performing BZ oscillations exhibit transitions between the two carrier types only for
certain $V_\mathrm{SD} = e E_D L$ when the phase $\phi$ fulfills the 
condition of Eq.~\eref{phasecond} as shown in the previous section. 
In consequence the current is strongly enhanced if this is fulfilled.
As shown in Fig.~\subref{figGrT}{a,b} the current peaks calculated by Eq.~\eref{currentEq} perfectly coincide
with the expected voltages (marked by vertical arrows) deduced by extracting the area $\mathcal A$ in momentum
space from the minibandes around the Fermi energy shown as shaded area in the inset of Fig.~\subref{figGrT}{a}.
Vice versa, the experimental observation of BZ peaks in the $I$-$V$ characteristics would allow for
`measuring' the miniband structure. 

A closer look at the transmission values $T(E, V_\mathrm{SD})$ [see inset Fig.~\subref{figGrT}{b}] reveals a
rhombic structure which features pronounced transmission maxima piled up at these particular values of
$V_\mathrm{SD}$ (dashed lines).
Since these maxima are present for various energies in the conductance window, the resulting current
is fairly independent of the exact Fermi energy [see Fig.~\subref{figGrT}{a}] and temperature
[see Fig.~\subref{figGrT}{b}].

\section{BZ oscillations and transport in HgTe-based superlattices}
\label{sec::HgTe}
%
%
\begin{figure}[tb]
\centering
\includegraphics[width=\stdwidth]{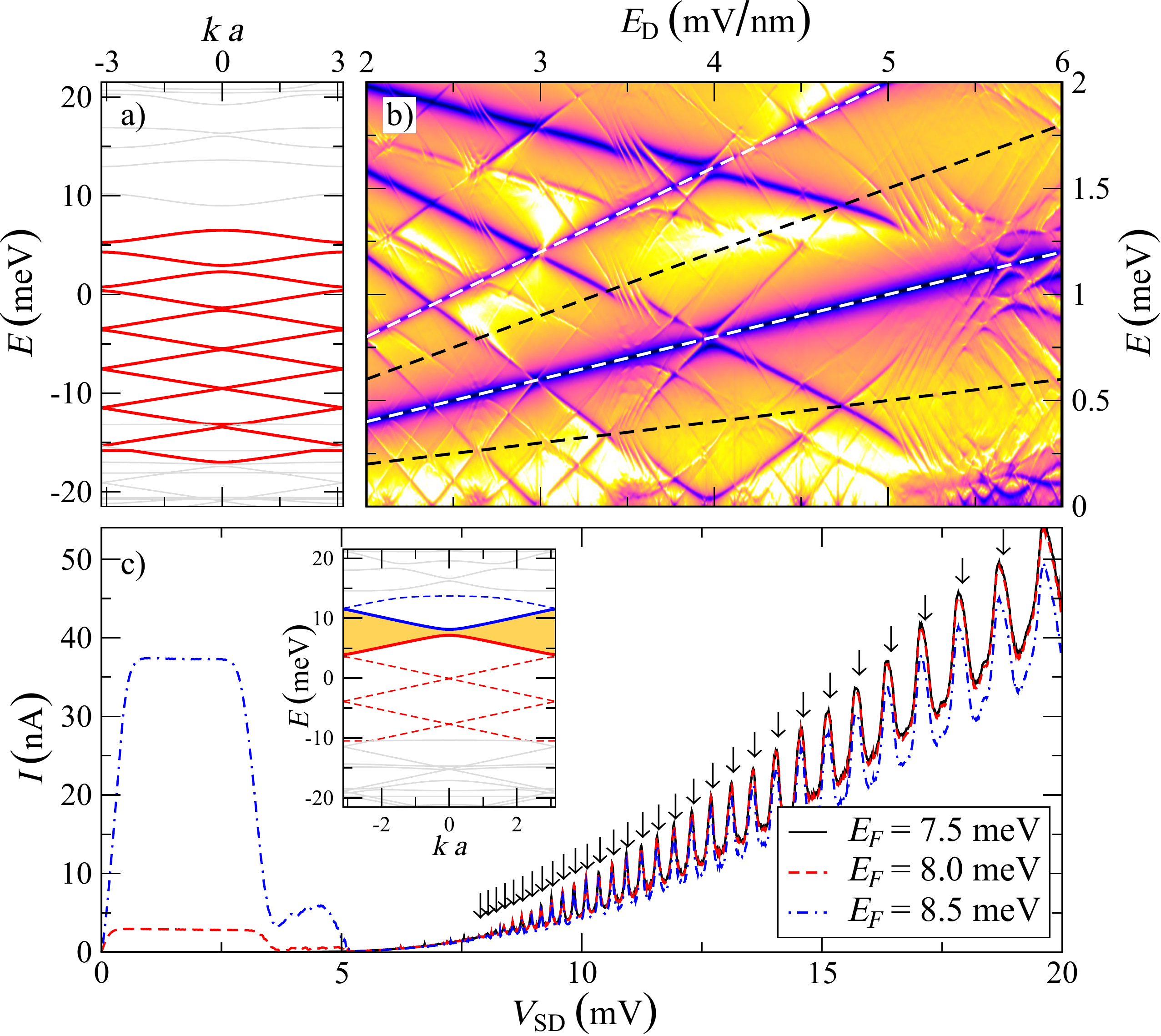}
\caption{(\emph{Color online})
Bloch and Bloch--Zener oscillations in spatially modulated two-dimensional HgTe nanoribbons.
\sub{a})
Bandstructure for a HgTe nanoribbon with periodically modulated width $W(x)$ (Eq.~\eref{width modulation}) ranging from 
$W_0=300~\mathrm{nm}$ to $W_1=50~\mathrm{nm}$ and periodicity $a=200~\mathrm{nm}$.
\sub{b})
Frequency spectrum $E=\hbar \omega$ of the wave packet center-of-mass motion as a function of drift field $E_D$.
Dashed lines indicate the frequencies of the Bloch oscillations. 
\sub{c})
$I$-$V_\mathrm{SD}$ characteristics of a nanoribbon with constant width $W=150~\mathrm{nm}$ and electrostatic
modulation $V(x) = V_0 \sin(2 \pi x / a)$. Small vertical arrows mark expected maxima from phase
condition~\eref{phasecond}. Inset: corresponding miniband structure.
\label{figHgTeT}
}
\end{figure}
%
%
A different setup featuring BZ oscillations can be created from a strip etched out of the two-dimensional
topological insulator based on mercury teluride (HgTe)~\cite{Koenig2007, Roth2009}.
We describe the electronic properties of the underlying HgTe heterostructure by 
the Hamiltonian~\cite{Bernevig2006}
%
%
\begin{equation}
H = \left ( \begin{matrix}
C_k + M_k & A k_+ &  0 & 0  \\
A k_- & C_k  - M_k & 0 & 0 \\
0 & 0 &  C_k  + M_k  & -A k_-\\
0 & 0 & -A k_+& C_k - M_k\\
\end{matrix}
\right )
\label{Hhgte}
\end{equation}
%
%
where $k_\pm = k_x \pm \ci k_y$, $\myvec k^2 = k_x^2 + k_y^2$, $C_k = -D \myvec k^2$ and $M_k = M - B \myvec k^2$.
We assume an HgTe/HgCdTe heterostructure with a quantum well width of $7.0~\mathrm{nm}$
featuring topological edge states and leading to material parameters
$A$, $B$, $D$, $M$ as typically used in literature~\cite{Koenig2008}.
As for the graphene nanoribbon we can create two different types of superlattices 
with a mass-like modulation and an electrostatic modulation.
For a HgTe strip the mass modulation can be achieved by modulating the width of the
ribbon, for example by
%
%
\begin{equation}
W(x) = \frac{W_0+W_1}{2}  - \frac{W_0-W_1}{2}  \sin\left( \frac{2 \pi x }{ a } \right)
\label{width modulation}
\end{equation}
%
%
where $W_0$ and $W_1$ are the maximum and minimum width and $a$ is the periodicity. 
The finite width of the HgTe nanoribbon allows for a hybridization of the edge states with
the same spin at the opposite boundaries  
leading to a small gap in the bandstructure~\cite{Zhou2008, krueckl2011}.
Accordingly, the modulation of the width corresponds to a modulation of the mass gap.
The resulting miniband structure from the two-dimensional system,
shown in Fig.~\subref{figHgTeT}{a}, is obtained numerically by Lanczos
diagonalization and exhibits various Landau-Zener anticrossings within the
bulk bandgap of HgTe which suggest BZ oscillations.

In order to study the electron dynamics we calculate the COM motion of Gaussian shaped edge-state wave packets.
Initially, the wave packet is localized on one edge and the direction of motion is determined by its spin.
The array of multiple constrictions enables tunneling between the edges.
As a consequence an inversion of the direction of motion is possible, leading to Bloch and
BZ oscillations.
As shown in Fig.~\subref{figHgTeT}{b}, the resulting frequency spectrum features the expected rhombic pattern
in between the frequencies of the conventional Bloch oscillations (white dashed lines).
Compared to the graphene system [see Fig.~\subref{figwpgraphene}{e}] we observe more complicated,
superimposed structures because of the whole sequence of multiple anticrossings in the band structure
that affect BZ oscillations.

As for graphene we further study the transport properties of HgTe strips of constant width and a periodically
modulated electrostatic potential resulting in a supercell bandstructure shown in the inset of
Fig.~\subref{figHgTeT}{c}.
The small gap between the electron and the hole states is attributed to the finite
ribbon width of $150~\mathrm{nm}$.
We chose the Fermi energy close to the band crossing of the topological edge states and calculate the current
using Eq.~\eref{currentEq}.
Besides a strong negative-differential conductance at lower bias we get
the signatures of BZ oscillations for
$V_\mathrm{SD} > 9~\mathrm{mV}$ as shown in Fig.~\subref{figHgTeT}{c}.
Similar to the calculations for the graphene superlattice the oscillations are independent of the exact
choice of the Fermi level.
The peak positions are in good accordance with the expected series of drift voltages
marked by arrows in Fig.~\subref{figHgTeT}{c} obtained from Eq.~\eref{phasecond},
where $\mathcal A$ is extracted from the bands around the
Fermi-energy shown as shaded area in the inset.

\section{Conclusion}
In this manuscript we showed that Bloch--Zener oscillations appear naturally in superlattices made of materials
with a Dirac-like spectrum, highlighting interference between electron and hole states.
The characteristics of these oscillations are explained by a one-dimensional model Hamiltonian and numerically
confirmed for realistic setups by means of wave packet simulations for graphene and topological
insulator ribbons.
Furthermore, we demonstrated that Bloch--Zener oscillations manifest themselves
as regular sequence of pronounced current peaks in quantum transport, besides
the well know negative differential conductance at low bias, a signature of
conventional Bloch oscillations.
The sequence of current peaks associated with the Bloch--Zener
oscillations are intimately linked to the underlying miniband structure.


We suggest transport measurements through graphene nanoribbons and HgTe strips as promising
experimental setups that feature Bloch--Zener oscillations.
For single layer graphene and topological insulators, the periodic electrostatic
potential can be imprinted by an array of top gates.
The gap between the electron and hole states can be tuned by the width of the
considered nanoribbons.
In case of bilayer graphene the gap can also be created via a potential 
difference in $z$-direction induced by top gating.
The calculations presented here have been performed for clean, disorder free and
coherent systems.
However, preliminary numerical calculations for graphene-based superlattices
with disorder indicate that Bloch--Zener oscillations are still visible 
if the mean-free-path exceeds several periods of the superlattice.
This is promising with respect to their experimental detection in sold-state
based samples.

We finally note that signatures of the Bloch--Zener oscillations presented have been recently
observed with ultracold, fermionic K-atoms due to the Dirac points with small mass gaps
emerging in tunable optical honeycomb lattices~\cite{Tarruell2012}.

{\em Note added in proof.} Recently, we became aware of
Ref. 47 where the Bloch-Zener oscillations of collective
excitations in narrow zigzag-shaped optical lattices is studied
theoretically.

\begin{acknowledgments}
This work is supported by Deutsche Forschungsgemeinschaft
(GRK 1570 and joined DFG-JST Forschergruppe Topological Electronics).
We thank T. Hartmann, F. Tkatschenko and D. Ryndyk for useful conversations.
\end{acknowledgments}

\bibliographystyle{prsty}

\end{document}